\documentclass[showpacs,preprintnumbers,amsmath,prl,
graphicx,amssymb,superscriptaddress,twocolumn]{revtex4}
\usepackage{graphicx}
\usepackage{dcolumn}
\usepackage{bm}

\renewcommand{\today}{\number\day\space\ifcase\month\or January\or 
 February\or March\or April\or May\or June\or July\or August\or 
 September\or October\or November\or December\fi\space\number\year}

\begin{document}
\title{Search for an Annual Modulation in a P-type Point Contact \\
Germanium Dark Matter Detector}

\newcommand{\pnnl}{Pacific Northwest National Laboratory, Richland,
WA 99352, USA}
\newcommand{\uc}{Kavli Institute for Cosmological Physics and Enrico
Fermi Institute, University of Chicago, Chicago, IL 60637, USA}
\newcommand{\canberra}{CANBERRA Industries, Meriden, CT 06450, USA}
\newcommand{\uw}{Center for Experimental Nuclear Physics and
Astrophysics and Department of Physics, University of Washington,
Seattle, WA 98195, USA}
\newcommand{\ornl}{Oak Ridge National Laboratory, Oak Ridge, TN 37831, USA}
\newcommand{\unc}{Department of Physics and Astronomy, University of 
North Carolina, NC 27599, USA}


\affiliation{\pnnl}
\affiliation{\uc}
\affiliation{\canberra}
\affiliation{\uw}
\affiliation{\ornl}
\affiliation{\unc}
														
\author{C.E.~Aalseth}\affiliation{\pnnl}
\author{P.S.~Barbeau}
\altaffiliation[Present address: ]{Department of Physics, Stanford 
University, Stanford, CA 94305, USA}
\affiliation{\uc}
\author{J.~Colaresi}\affiliation{\canberra}
\author{J.I.~Collar}
\email[Contact author: ]{collar@uchicago.edu}
\affiliation{\uc}
\author{J.~Diaz Leon}\affiliation{\uw}
\author{J.E.~Fast}\affiliation{\pnnl}
\author{N.~Fields}\affiliation{\uc}
\author{T.W.~Hossbach}\affiliation{\pnnl}\affiliation{\uc}
\author{M.E.~Keillor}\affiliation{\pnnl}
\author{J.D.~Kephart}\affiliation{\pnnl}
\author{A.~Knecht}\affiliation{\uw}
\author{M.G.~Marino}
\altaffiliation[Present address: ]{Physics Department, Technische 
Universit\"{a}t M\"{u}nchen, Munich, Germany}
\affiliation{\uw}
\author{H.S.~Miley}\affiliation{\pnnl}
\author{M.L.~Miller}\affiliation{\uw}
\author{J.L.~Orrell}\affiliation{\pnnl}
\author{D.C. Radford}\affiliation{\ornl} 
\author{J.F.~Wilkerson}\affiliation{\unc}
\author{K.M.~Yocum}\affiliation{\canberra}

\collaboration{CoGeNT Collaboration}
\noaffiliation
```

\begin{abstract}
Fifteen months of cumulative CoGeNT 
data are examined for indications of an annual modulation, a predicted signature 
of Weakly Interacting Massive Particle (WIMP) interactions. Presently available data 
support the presence of a modulated component of unknown origin,
with parameters {\it prima facie} compatible with a 
galactic halo composed of light-mass WIMPs. Unoptimized estimators yield a statistical significance for a modulation of  $\sim\!2.8 \sigma$, limited by the short exposure.
\end{abstract}

\maketitle

CoGeNT employs P-type Point Contact (PPC) germanium detectors \cite{ourprl2,jcap,ourprl1} to 
explore light-mass WIMP dark matter models. By virtue of their 
low electronic noise, PPCs are ideal for searches in the mass range m$_{\chi}<10$ GeV/c$^{2}$. Their ability to reject  
background events taking place on detector surfaces is
described in \cite{ourprl2}. Prompted by the interruption in data-taking imposed by a 
recent fire in the access shaft to the Soudan Underground Laboratory 
(SUL), we have examined existing CoGeNT data for the presence of an 
annual modulation in the WIMP interaction rate 
\cite{andrzej}. The characteristics of 
this detector, data analysis and background interpretations are treated in a recent Letter \cite{ourprl2}, to which the reader 
is referred.

Underground installation of this PPC at SUL took place on August 21, 
2009. 
Following an upgrade to the data acquisition to 
allow discrimination of surface events, 
data-taking started on December 4th, a date close 
to the minimum in rate expected from the annual modulation effect 
\cite{andrzej}.
Data-taking was interrupted exclusively over the 
periods of February 9-15, March 15-20, and October 5-7 of 2010, for 
inspection of a higher-energy region of the spectrum and two
planned general power outages at SUL. The dataset presented here ends on 
March 6 of 2011, spanning 458 days, of which 442 were live. On this date 
the detector was stopped for another planned outage. 
Outage-related computer problems delayed normal operation until two days before the 
fire (March 17, 2011). 

Surface background- and microphonic-rejection cuts 
on these data are as in \cite{ourprl2}, and 
constant in time. An alternative analysis \cite{mike} 
results into  
very similar conclusions, to be treated elsewhere \cite{inprep}.
In the following discussion, the astrophysical 
halo parameters in \cite{CDMSsoudan,lastXENON} are used.

Fig.\ 1 (top) displays spectral peaks appearing at the K-shell binding energy of  
daughters of cosmogenically-activated radioisotopes decaying via 
electron capture (EC) \cite{ourprl2}. 
The bottom panel in the same figure zooms in to the lower 
energy region of the spectrum, down to the $\sim$0.5 keV$_{ee}$ (keV 
electron equivalent or ionization energy) threshold. A fit to the 
evolution of the K-shell peaks returns excellent agreement with the expected half-lives and allows a prediction for 
the initial abundance of these radioisotopes with individual 
uncertainty of O(10)\%. 
The ratio between L-shell and K-shell EC is well-defined for these 
isotopes, both 
theoretically and experimentally \cite{bahcall}. It can be used to 
generate a prediction, devoid of any free parameters,
for the intensity of the L-shell peaks 
in this lower-energy spectrum (Gaussians in Fig.\ 1, bottom). 
After subtraction of this predicted L-shell EC contribution and of a 
constant spectral component, and following the correction for the 
combined trigger and software-cut efficiency \cite{ourprl2}, an 
exponential-like irreducible background of events taking place in the 
bulk of the crystal is observed (Fig.\ 1, inset). These events are 
individually distinct from any identifiable source of noise, display the characteristics (rise and decay time) of radiation-induced pulses, and have a uniform diurnal distribution.  
Predicted signals from 
example combinations of light-WIMP masses and spin-independent 
WIMP-nuclei couplings are also shown as a reference.

\begin{figure}
\includegraphics[width=7.cm]{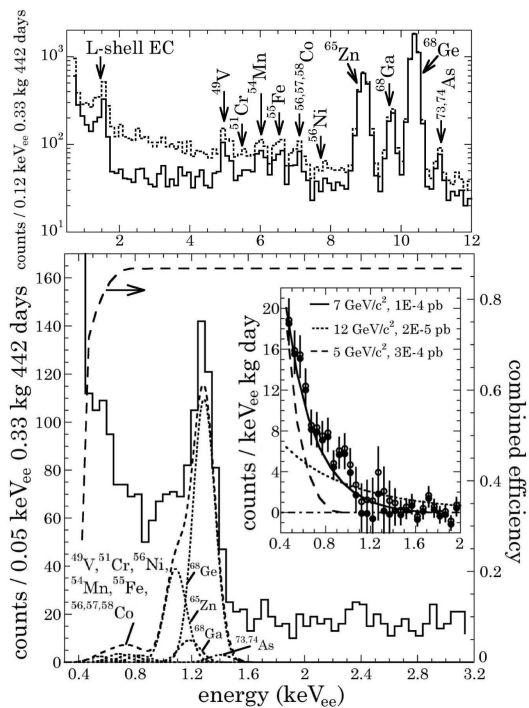}
\caption{\label{fig:epsart}{\it Top:} Uncorrected (i.e., prior to threshold 
efficiency correction) spectrum displaying 
all expected K-shell EC cosmogenic peak positions. The dotted histogram shows 
the spectrum before rejection of surface background events. {\it 
Bottom:} Uncorrected low-energy spectrum following removal of surface events. 
Dotted Gaussian peaks show the predicted L-shell EC contribution, 
devoid of any free parameters (see text). A dashed line traces their envelope.  
A second dashed line indicates the combined threshold  
efficiency (trigger + software cuts)  \protect\cite{ourprl2}, an arrow pointing from it to the right scale. 
{\it Inset:} Spectra corrected by this efficiency 
and stripped of L-shell contribution and flat background component. 
Examples of light WIMP signals are overlapped on it (see text). 
}
\end{figure}

Black dots in the inset of Fig.\ 1 represent the irreducible spectrum obtained 
by stripping of the L-shell predictions and the flat background level 
in the region $\sim$2-4.5 keV$_{ee}$. Unfilled circles are obtained 
when a free overall normalization factor multiplies the 
envelope of the individual L-shell predictions in a background model 
containing this envelope, an exponential and a constant background. The 
resulting best-fit indicates a L-shell contribution just 10\% short 
of the nominal prediction, well within its uncertainty. Fig.\ 2 shows the 
region of interest (ROI) obtained when these irreducible spectra are 
fitted by a sample model containing signals from WIMPs of mass 
m$_{\chi}$ and spin-independent coupling $\sigma_{SI}$, and a free 
exponential background. As in 
\cite{ourprl2}, this ROI is defined  
by the upper and lower 90\% C.L. 
intervals for the best-fit $\sigma_{SI}$, whenever the lower interval is incompatible with a null value. 
This ROI is meant to direct the eye to the region of 
parameter space where the hypothesis of a WIMP signal dominating the irreducible
background events fares best, but it does not include astrophysical or other uncertainties listed next. Reasonable uncertainties in the germanium 
quenching factor employed (Fig.\ 4 in \cite{jcap}, \cite{phil}) can shift this ROI 
by $\sim\pm$1 GeV/$c^{2}$. The present uncertainty in the fiducial bulk 
volume of this detector is O(10)\% \cite{ourprl2}. Departures from the assumption of a constant background in the model above can also displace this region. A modest contamination of the spectrum by surface events next to threshold \cite{ourprl2,inprep} would shift this ROI to slightly higher values of m$_{\chi}$ and lower $\sigma_{SI}$.
The additional exposure collected since 
\cite{ourprl2} results in a much reduced CoGeNT ROI, 
one in the immediate vicinity of the parameter space compatible with the annual 
modulation effect observed by DAMA/LIBRA \cite{prddan,dama}. This 
region of $\sigma_{SI}$, m$_{\chi}$ space is populated by the predictions 
of several particle phenomenologies. The reader is directed to  
references in \cite{ourprl2} and recent literature for examples. The same region 
has received recent attention within the context of dark matter 
annihilation signatures at the center of our galaxy, and anomalies in 
accelerator experiments \cite{other}.
Fig.\ 2 also displays limits from 
other searches, a subject treated again below.

\begin{figure}
\includegraphics[width=7.cm]{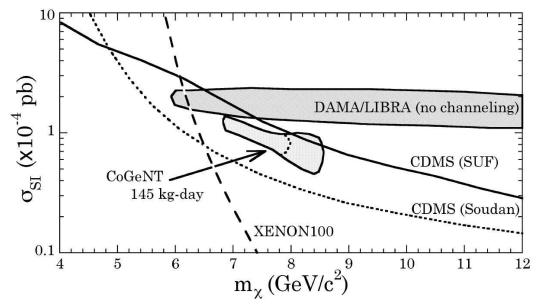}
\caption{\label{fig:epsart}ROI extracted from 
the irreducible spectra in Fig.\ 1 (inset) under consideration of a light-WIMP hypothesis. A small dotted 
line bisects it, approximately separating the domains favored by the black 
dot (left) or unfilled circle (right) spectra in Fig.\ 1.  ROI definition and uncertainties able to shift it  are described in the text. The DAMA/LIBRA ROI includes 
present uncertainties in its position
\protect\cite{prddan}, with the exception of ion channeling 
\protect\cite{channeling},  
conservatively assumed to be absent. Solid and dotted lines are CDMS limits from 
\protect\cite{CDMSSUF} and \protect\cite{CDMSsoudan}, respectively. A dashed line corresponds
to recent XENON100 claims \protect\cite{lastXENON}. Uncertainties in these constraints and those by XENON10
\protect\cite{peter} are examined in \protect\cite{cdmscrit,comp}.}
\end{figure}

A search for a WIMP-induced annual modulation in dark matter detector 
data requires an exceptional low-energy stability in the device. Fig.\ 3 shows 
that these conditions are  present for CoGeNT. The top panel 
displays daily averages in the detector electronic noise. 
Excessive excursions in this parameter would affect the stability of 
the detector threshold. These are not observed. Precautions are taken to 
ensure that this noise is as stable as possible: for instance, by
automatically refilling the detector liquid nitrogen Dewar every 48h, 
the crystal temperature and its associated leakage 
current are held as constant as possible. The second panel shows the 
stability of the trigger threshold, derived from the difference between the daily average 
baseline DC level in the triggering channel and a constant (digitally fixed) discriminator level. The small excursions observed correspond to a temperature drift in the 
digitizers (NI 5102) and shaping amplifier (Ortec 672) 
of $\sim1^{\circ}$C. These small instabilities do not result in any
minor smearing of the energy resolution, 
given that the amplitude of  
each event is referenced to its individual 
pre-trigger DC level. The effect of this small baseline drift should 
instead be envisioned as a maximum shift of the threshold efficiency curve in Fig.\ 1 
by about $\pm$10 eV. The third panel shows the calculation of by how 
much such a shift can affect the counting rate in the region 0.5-0.9 
keV$_{ee}$. This calculation includes the exponential spectral shape 
observed there. The correction is referenced to the date of the 
threshold efficiency calibration employed (small arrow in Fig.\ 3) 
and found to be negligible at less than 0.1\%. This correction would be larger for 
events below 0.5 keV$_{ee}$, not considered here, and even smaller 
for count rates in broader energy regions starting at 0.5 
keV$_{ee}$. 
The fourth panel indicates the magnitude of the correction required 
to account for the exponential decay of L-shell EC radioisotopes,
prior to an annual modulation analysis. 
This correction affects the 0.4-1.6 keV$_{ee}$
region (Fig.\ 1), where a light-WIMP can express a modulated signature. The 
uncertainties in this correction, indicated in Fig.\ 3 in parentheses,  
are modest even at the present exposure. A direct comparison of 
these predictions with the low-energy spectrum, as done in Fig.\ 1, 
demonstrates that they are robust.

\begin{figure}
\includegraphics[width=7. cm]{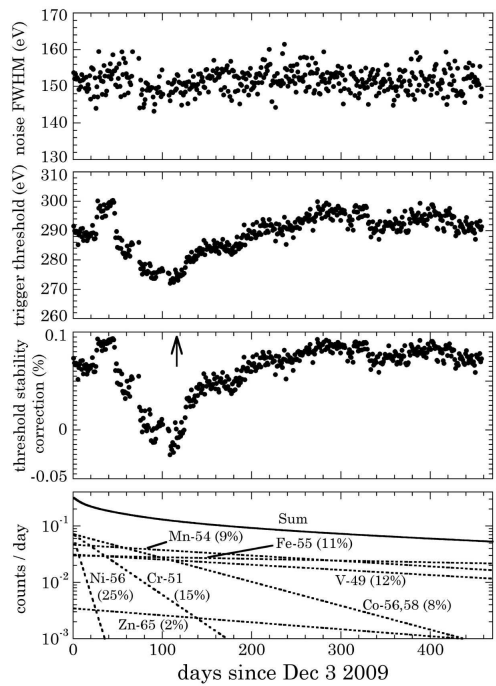}
\caption{\label{fig:epsart}Assessment of the stability of the 
CoGeNT PPC at Soudan (see text). First panel: daily average in detector electronic noise 
(shaping time 10 $\mu$s). Second panel: stability of the trigger threshold 
level. Third panel: negligible correction to the counting rate in the region 0.5-0.9 
keV$_{ee}$ induced by it. Fourth panel: expected counting 
rate in this same region originating in L-shell EC. The observed 
stability augurs well for WIMP modulation searches using next-generation 
PPCs like those planned for the upcoming expansion of CoGeNT (C-4), 
MAJORANA, GERDA, and CDEX. 
}
\end{figure}

Fig.\ 4 shows the temporal rate behavior in several spectral regions 
following the correction for L-shell EC activity, when applicable. 
Even with the present limited exposure, a noticeable annual modulation can be
observed in the energy region encompassing the exponential rise in 
irreducible background in Fig.\ 1. The statistical significance for 
this modulation is presently maximal for an energy bin ranging from 
threshold (0.5 keV$_{ee}$) to an upper bracket anywhere in the interval 
$\sim$2.0-3.0 keV$_{ee}$
(the amplitude of a WIMP-induced modulation 
is expected to be maximal towards its spectral endpoint 
\cite{pfsmith,josh}). 
No indication of a modulation is observed above 
this energy, 
nor for events rejected as surface backgrounds in the same spectral region 
where it is the largest for bulk events (bottom panel in Fig.\ 4). The 
``internal clock'' provided by the most intense K-shell peak at 
10.37 keV$_{ee}$ exhibits no modulated deviation from an exponential 
decay, nor any significant time-dependent changes in mean energy.

{\it Ad hoc} methods able to extricate a modulated component with maximal sensitivity \cite{josh}
have not been attempted. A few considerably less sensitive indicators of 
modulation significance are 
presently offered. For instance, for the region 0.5-3.0 keV$_{ee}$ and (unoptimized) binning 
depicted in Fig.\ 4, a straightforward analysis reveals a reduced 
chi-square $\chi^{2}$/{\frenchspacing d.o.f.}=7.8/12 (80\% C.L. for acceptance) 
for the best-fit 
modulation. The null hypothesis (absence of a modulation) fares 
considerably worse at $\chi^{2}$/{\frenchspacing d.o.f.}=20.3/15 (84\% C.L. for rejection). The likelihood ratio test indicates that the modulation hypothesis is preferred over the null hypothesis at  99.4\% C.L. (2.8$\sigma$).
Intriguingly, the best-fit values for the three modulation parameters
(16.6$\pm$3.8\% modulation amplitude, period 347$\pm$29 
days, minimum in Oct. 16$\pm$12d) do not fall far from the predictions 
provided by the WIMP hypothesis (a calculable amplitude, a yearly 
period and minimum amplitude in late Nov.\ to early Dec. \cite{kuhlen}). The 
most uncertain of these predictions \cite{kuhlen,nealmod}, the amplitude, is 
derived as in \cite{pfsmith}, including 
the dependence on astrophysical parameters, target, 
threshold, WIMP mass, etc. For  m$_{\chi}$=7 GeV/c$^{2}$ an 
expected value of 
12.8\% is obtained (Fig.\ 4). While the apparent DAMA/LIBRA 
modulation is weaker 
($\sim$2\% of the low-energy rate), its enhancement for CoGeNT is expected in most light-WIMP 
scenarios \cite{nealmod,danmod}. 
If these predictions are accepted, the 
Monte Carlo probability of better simultaneous 
agreement with them than that provided by the best-fit values above, 
is small at 0.7\%, for simulated random fluctuations around the 0.5-3.0 keV$_{ee}$ mean rate. 

In addition to the basic estimators presented here, time-stamped 
CoGeNT data are available by request. These can be used not only for 
alternative analyses of significance, etc.,
but also to investigate non-cosmological effects that 
might generate this modulation and by extension that observed by 
DAMA/LIBRA.  In this respect, the muon flux at SUL varies seasonally by $\pm$2\%, and radon levels by a factor $\sim$4 \cite{goodm}. Muon-coincident events constitute a few percent of the low-energy spectrum \cite{ourprl2}, limiting a muon-induced modulated amplitude to $<<$1\% \cite{inprep}. 
Rejection of veto-coincident events does not alter the observed modulation. 
Radon displacement via pressurized LN boil-off gas is continuously maintained at 2 l/min within an aluminum shell encasing the lead shielding \cite{www}. A radon-induced modulation would be expected to affect a much broader spectral region than observed \cite{radon}.

\begin{figure}
\includegraphics[width=7.cm]{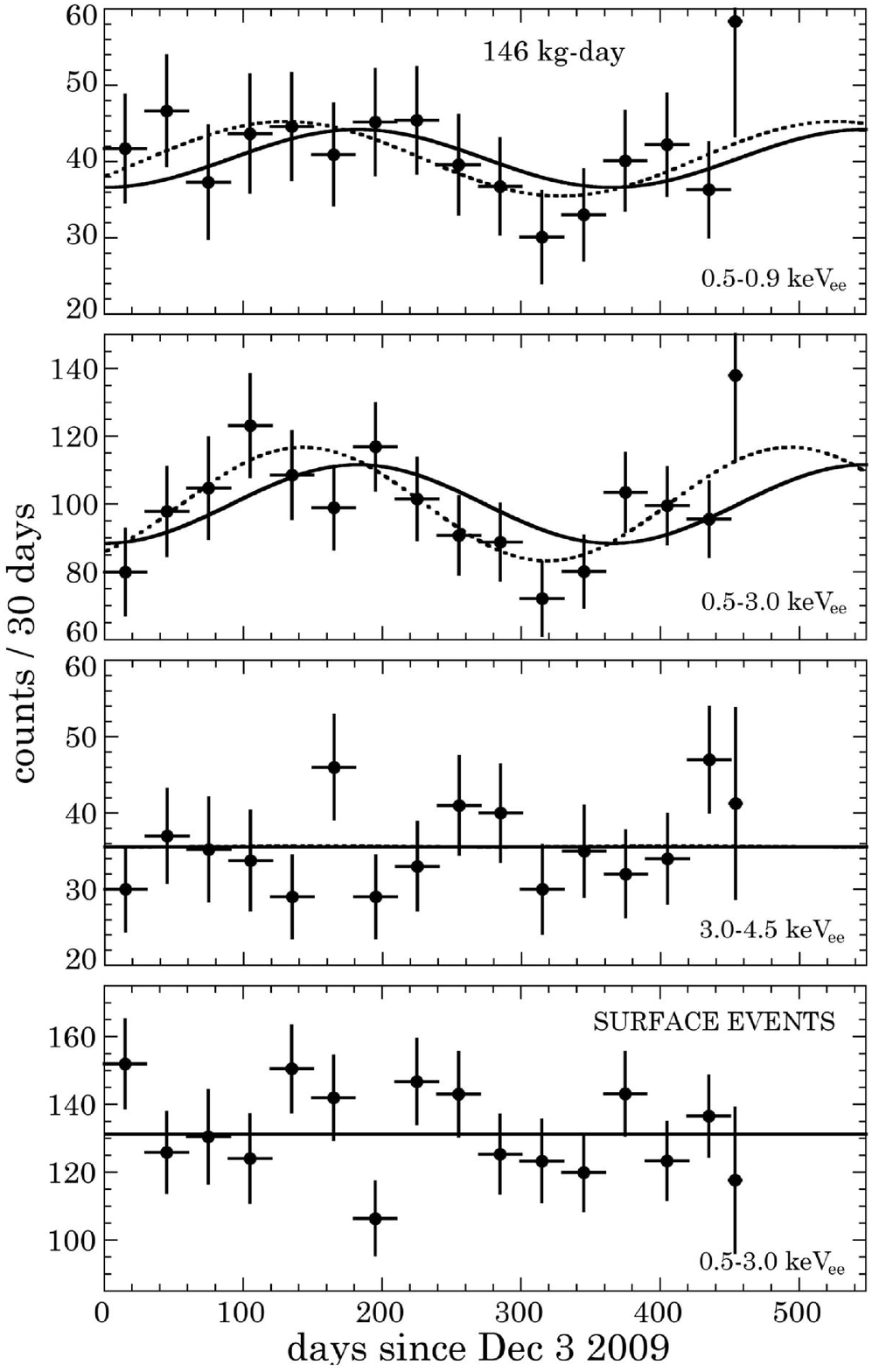}
\caption{\label{fig:epsart}Rate vs.\ time in several energy regions (the last bin spans 8 
days). A dotted line denotes the best-fit modulation.
A solid line indicates a prediction for a 7 GeV/c$^{2}$ WIMP in a galactic halo with Maxwellian velocity distribution. Background contamination and/or a non-Maxwellian halo can shift the amplitude of this nominal modulation (see text). Dotted and solid lines overlap for the bottom panels.   
}
\end{figure}

The CDMS collaboration 
has recently claimed \cite{CDMSsoudan} to exclude a light-WIMP 
interpretation of CoGeNT and 
DAMA/LIBRA observations. Uncertainties affecting this claim are discussed in \cite{cdmscrit,cdmsv3}.
Observations from 
XENON10 \cite{peter} and XENON100 \cite{lastXENON} have been used 
to claim a similar rejection of light-WIMP scenarios. 
Uncertainties affecting these searches are examined in \cite{comp}. 

In conclusion, presently available CoGeNT data favor the presence of 
an annual modulation in the low-energy spectral rate, for events taking place in 
the bulk of the detector only. While its origin is presently unknown, the spectral and temporal information 
are {\it prima facie} congruent when the WIMP hypothesis is examined: in particular, 
the WIMP mass region most favored by a spectral analysis (Fig.\ 2) generates 
predictions for the modulated amplitude in agreement with 
observations, modulo the dependence of this assertion on 
the choice of astrophysical parameters  and halo velocity distribution \cite{kuhlen,nealmod,danmod,nonmax}. 
 
\section{ACKNOWLEDGMENTS}
Work sponsored by NSF grants PHY-0653605 and PHY-1003940, The Kavli 
Foundation and  PNNL
LDRD program. N.F. and T.W.H. are supported by the DOE/NNSA SSGF program and the National Consortium for MASINT
Research, respectively. We owe much gratitude to SUL personnel for their assistance and to D. Hooper and N. Weiner 
for many useful exchanges.

\end{document}